\documentclass[prl,aps,twocolumn,showpacs,floatfix]{revtex4}
\usepackage{graphicx}
\usepackage{amsmath}

\begin{document}

\title{Resistivity of inhomogeneous quantum wires}

\smallskip

\author{J\'er\^ome Rech and K. A. Matveev} \affiliation{Materials Science
  Division, Argonne National Laboratory, Argonne, Illinois 60439, USA}
\date{\today} 
\begin{abstract}
  
  We study the effect of electron-electron interactions on the transport
  in an inhomogeneous quantum wire. We show that contrary to the
  well-known Luttinger liquid result, non-uniform interactions contribute
  substantially to the resistance of the wire.  In the regime of weakly
  interacting electrons and moderately low temperatures we find a linear
  in $T$ resistivity induced by the interactions.  We then use the
  bosonization technique to generalize this result to the case of
  arbitrarily strong interactions.

\end{abstract}
\maketitle

Since the first measurements of quantized dc conductance in quantum wires
\cite{quantized}, transport properties of these systems have generated a
lot of interest. Theoretically, such one-dimensional conductors cannot
be described by the conventional Fermi liquid theory, but rather form
a qualitatively different state known as the Luttinger liquid
\cite{LL-th}.  Recently several characteristic signatures of the
Luttinger liquid state have been reported experimentally in quantum
wires \cite{nanotube,LL-exp}.  Perhaps even more interestingly, a
number of experiments show anomalies in the transport properties of
these systems in the form of a small structure below the first plateau
of quantized conductance
\cite{thomas2,reilly,thomas1,kristensen,cronenwett,depicciotto}, which 
are not expected in the Luttinger liquid theory
\cite{FLleads}. The so-called ``0.7-structure,'' which develops at
finite temperature in low-density wires, is an example of such
deviations from perfect quantization
\cite{thomas1,kristensen,cronenwett,depicciotto}. While its precise
origin still remains unclear, this feature is most likely related to
correlations between electrons, which initiated various attempts to
study the effect of interactions on the electronic transport in these
devices \cite{7theory, Kostya, sushkov}.

A number of recent theory papers \cite{sushkov, FLleads} studied the model
of a quantum wire device in which interactions are present only in a small
region of a one-dimensional electron system between two non-interacting
leads.  If the size of the interacting region does not significantly
exceed the Fermi wavelength of the electrons in the wire, the interactions
give rise to backscattering of either single electrons or pairs, resulting
in significant corrections to the quantized conductance \cite{sushkov}.
On the other hand, if the interaction strength varies smoothly over a long
distance, such backscattering processes are expected to be exponentially
weak and can be neglected.  In this regime a model of non-uniform
Luttinger liquid, with parameters gradually varying as a function of
position is appropriate.  Studies of such a model found no correction to
the quantized dc conductance of the wire \cite{FLleads}.  It is thus
natural to conclude that inhomogeneities of interacting quantum wires at
large scales $d\gg k_F^{-1}$ do not affect the dc transport beyond the
exponentially small backscattering corrections.

In this paper we show that even at $k_F d\gg 1$, when the backscattering
processes \cite{sushkov} can be ignored, the inhomogeneity of the
interaction strength in the wire gives rise to a finite resistivity at
non-zero temperature.

We start by considering an infinite one-dimensional system of weakly
interacting spinless electrons, with quadratic dispersion $\epsilon_p =
p^2/2m$. In this simple model, the electron density $n$ is assumed to be
uniform, but the strength of the electron-electron interactions varies
along the wire.  We describe these inhomogeneous interactions by the
potential
\begin{equation}
\label{interaction}
{\cal V}(x,y) = V(x-y) ~ \eta \left( \frac{x+y}{2} \right).
\end{equation}
Here $V(x-y)$ is the conventional electron-electron repulsive interaction.
Coulombic in nature, it is screened by the nearby gates, and for
simplicity we will treat it as a short-range interaction. The
non-uniformity of the system is then encoded in the dimensionless function
$\eta$, which varies at a length scale $d$, large compared with both the
Fermi wavelength and the range of the interaction potential $V(x-y)$.

In order to compute the resistance of the wire, we enforce a dc current
$I$ to flow through the system. The electrons in the wire then acquire a
drift velocity proportional to this current: $v_d = I/n e$.  In the
reference frame moving along the wire with velocity $v_d$ the electronic
subsystem is in equilibrium, as pointed out by Pustilnik \emph{et~al.}
\cite{drag} in the context of Coulomb drag between two parallel wires.
This equilibrium is characterized by a Fermi energy $\epsilon_F$ and a
temperature $T$.

When viewed in the stationary reference frame, where the electric current
does not vanish, the electrons are no longer in thermodynamic equilibrium.
In particular, their occupation probabilities cannot, in general, be
expressed as a Fermi function of the energy.  However, at $T\ll
\epsilon_F$ the occupation probabilities of the left- and right-moving
states near the Fermi level can still be approximated by Fermi
functions, albeit with two different temperatures, $T_L$ and $T_R$.
To show that, we note that the electron energy $\epsilon_p$ changes to
a different value $\tilde\epsilon_p$ in the stationary frame.
Considering a state $p$ near the right Fermi point $p_F$, to first
order in $p-p_F$, we have
\[
  \frac{\epsilon_p-\epsilon_F}{\tilde\epsilon_p-\tilde\epsilon_F}
  = \frac{v_F(p-p_F)}{\tilde v_F(\tilde p-\tilde p_F)}
  = \frac{v_F}{\tilde v_F},
\]
where in the stationary frame $\tilde v_F = v_F + v_d$, $\tilde p =
p+mv_d$, $\tilde p_F=p_F+mv_d$.  Thus, the occupation probability of this
state can be expressed in terms of its energy $\tilde \epsilon_p$ in the
stationary frame as
\[
  \frac{1}{e^{(\epsilon_p-\epsilon_F)/T}+1}=
  \frac{1}{e^{(\tilde\epsilon_p-\tilde\epsilon_F)/T_R}+1},
\]
where $T_R=T\tilde v_F/v_F=T(1+v_d/v_F)$.  Similarly, for the electrons
near the left Fermi point $-p_F$ one finds the occupation probability
given by the Fermi function with the effective temperature
$T_L=T(1-v_d/v_F)$.

\begin{figure}[tb]
\begin{center}
\includegraphics[width=82mm]{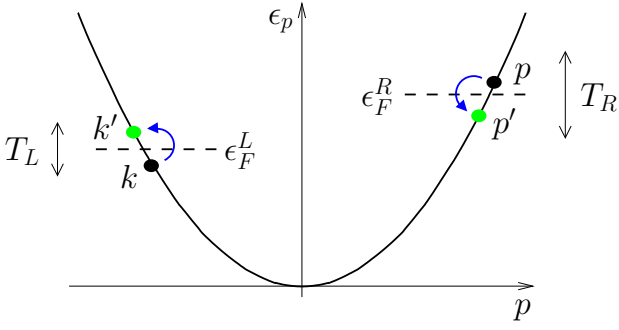} 
\end{center}
\vspace{-15pt} 
\caption{
        Electronic dispersion in the stationary frame, with an example
        of non momentum-conserving process. The right- and left-movers
        Fermi energies read $\epsilon_F^{R,L} = \epsilon_F (1
        \pm 2 v_d/v_F)$. }
\label{fig1}
\end{figure}

One may expect that the electron-electron interactions induce
thermalization between right- and left-movers through two-particle
scattering processes like the one shown in Fig.~\ref{fig1}. In a uniform
system, two-particle processes cannot lead to thermalization
\cite{3electron} because of the conservation of both energy and momentum:
the interactions either exchange the momenta of the two electrons or leave
them unchanged.  However, in our model the non-uniformity of the
interaction potential (\ref{interaction}) breaks the translational
invariance of the system, and allows for two-particle scattering processes
that conserve energy but not momentum.

A typical two-particle process shown in Fig.~\ref{fig1} describes the
scattering from a state with momenta $(p,k)$ to $(p^{\prime},k^{\prime})$,
and is accompanied by an overall loss of momentum.  Since this process
involves a transfer of energy from the ``warmer'' right-moving branch to
the ``colder'' left-moving one, it is expected to occur more frequently
than the inverse process $(p^{\prime},k^{\prime}) \to (p,k)$, so that on
average, the two subsystems lose more momentum than they gain.  Note that
unlike Ref. \onlinecite{sushkov}, the typical change of momentum for the
processes shown in Fig.~\ref{fig1} is small compared to the Fermi
momentum, and the rate of such processes will not become exponentially
small at $k_F d\gg1$.

The decrease in momentum can be interpreted as a result of a damping force
acting on the electrons. To maintain constant current, it has to be
balanced by a driving force, which stems from a local electric field,
generated as a response of the system to the current bias \cite{drag}.
Since the damping force is proportional to the temperature difference
between the two subsystems, $T_R-T_L\propto I$, this local electric field
is proportional to the applied current, and the proportionality
coefficient is defined as the resistivity.

Let us compute the resistivity at temperatures $T\gg \hbar v_F/d$.  We
isolate a small segment of wire taken at position $x$, with the length
$\Delta x$ in the range $\hbar v_F/T \ll \Delta x \ll d$. The driving
force $e E(x) n\Delta x$ acting on this segment of wire as a result of
the local electric field $E(x)=\rho(x) I$ compensates for the damping
force $\Delta F$ due to the interactions, so that the resistivity can
be written as
\begin{equation} \label{defrho}
\rho(x) = - \frac{\Delta F}{e n I \Delta x}.
\end{equation}
We compute the damping force as the change in momentum per unit
time, using the Fermi golden rule
\begin{eqnarray} 
\Delta F & = & \frac{2 \pi}{\hbar} \sum_{p,k,p^{\prime},k^{\prime}}  \left|V_{pk;p^{\prime}k^{\prime}}  \right|^2  \delta(\epsilon_{p}+\epsilon_{k}-\epsilon_{p^{\prime}}-\epsilon_{k^{\prime}}) \nonumber \\
& & \times \left( p^{\prime} + k^{\prime} -p-k \right) f_p^R f_k^L
(1-f_{p^{\prime}}^R) (1-f_{k^{\prime}}^L).
\label{pdot}
\end{eqnarray}
Here $V_{pk;p^{\prime}k^{\prime}}$ is the matrix element of the
interaction potential for scattering from an initial state $(p,k)$ to
a final state $(p^{\prime},k^{\prime})$, as shown in Fig.~\ref{fig1},
the occupation numbers $f^{R,L}$ are given by the Fermi-Dirac
distribution evaluated with the appropriate temperatures $T_{R,L}$.
One can easily check that at $T_R=T_L$ expression (\ref{pdot})
vanishes.  Then in the linear response regime, one can expand the
occupation numbers $f^R$ and $f^L$ to first order in $T_R-T_L\propto
I$, and find that $\Delta F$ is proportional to the applied current.

To first order in the interaction potential, the matrix element
$V_{pk;p^{\prime}k^{\prime}}$ is given by:
\begin{equation} \label{matrixelt}
V_{pk;p^{\prime}k^{\prime}} = \int_x^{x+\Delta x} dy e^{i\frac{
p^{\prime} + k^{\prime} - p - k}{\hbar} y} \left[ V(0)-V(2k_F)
\right]\eta(y),
\end{equation}
where $V(0)$ and $V(2k_F)$ are respectively the zero and $2 k_F$ Fourier
components of the interaction potential introduced in
Eq.~(\ref{interaction}). Substituting Eq.~(\ref{matrixelt}) into
Eq.~(\ref{pdot}), one readily sees that a constant value of $\eta$
enforces the conservation of momentum and leads to a vanishing result. The
dominant non-vanishing part thus involves the gradient of $\eta$, and
contributes to $\Delta F$ as $(\partial_x \eta)^2$.

Performing the remaining momentum summations, the resistivity evaluated to
second order in the interaction then takes the form
\begin{equation}
\rho(x)  =  \frac{h}{64 e^2} \frac{T}{n \epsilon_F} \left( \frac{V(0)-V(2k_F)}{\pi \hbar v_F}\right)^2  \left( \frac{\partial \eta(x)}{\partial x} \right)^2 .
\label{rhoresult}
\end{equation}
This result was obtained in the regime of temperatures $T \gg \hbar
v_F/d$, for which the position integrals coming from the matrix
element (\ref{matrixelt}) could be easily simplified.

Our method provides a clear physical picture of the origin of the
resistivity, in the simple case of weakly interacting spinless fermions.
We now turn to the case of arbitrarily strong interactions, where we
derive the expression for the resistivity using a bosonized Hamiltonian.
In addition, we account for electron spins and allow for a non-uniform
density $n(x)$, as a result of the surrounding gates and impurities in the
substrate.

Following Ref.~\onlinecite{FLleads}, we generalize the Tomonaga-Luttinger
model of interacting one-dimensional electron systems to account for
inhomogeneities by allowing for position dependence of the
Luttinger-liquid parameters.  In the case of electrons with spins, this
procedure yields
\begin{subequations}
\label{bosonized_Hamiltonian}
\begin{eqnarray}
H & = & H_{\rho} + H_{\sigma} \\ 
H_{\rho} & = & \int dx ~\frac{\hbar u_{\rho}(x)}{2 \pi} 
\left[ K_{\rho}(x) \left(\partial_x \theta_{\rho} \right)^2 
+ \frac{\left(\partial_x \phi_{\rho} \right)^2}{K_{\rho}(x)} \right] 
\label{bos-charge} \\ 
H_{\sigma}& = & \int dx ~\frac{\hbar u_{\sigma}(x)}{2 \pi} 
\left[ K_{\sigma}(x) \left(\partial_x \theta_{\sigma} \right)^2 
+ \frac{ \left(\partial_x \phi_{\sigma} \right)^2}{K_{\sigma}(x)}
\right] \nonumber \\ 
& & + \int dx ~ \frac{2 g_{\sigma} (x)}
{\left[2 \pi\alpha(x)\right]^2} \cos \left( 2\sqrt{2}
\phi_{\sigma}\right). 
\label{bosonized}
\end{eqnarray}
\end{subequations}
Here the short distance cutoff $\alpha$ is assumed to be a function of
$x$.  In the limit of a homogeneous system, $k_F d\to\infty$, the
coupling constant $g_{\sigma}$ renormalizes to zero at large length
scales; at the same time, the parameter $K_{\sigma}$ approaches unity
as $K_{\sigma}=1+g_{\sigma}/2\pi \hbar u_{\sigma}$, as required by the
SU(2) symmetry \cite{LL-th}.  We assume that $k_Fd$ is sufficiently
large for the system to be near this limit at every point $x$, i.e.,
$g_{\sigma}(x)/\hbar u_{\sigma}(x)\ll1$.  Previous works on the
inhomogeneous Luttinger liquid model were either restricted to
spinless electrons \cite{FLleads}, or discarded \cite{ITLL} the cosine
term in Eq.~(\ref{bosonized}), invoking the irrelevance of
$g_{\sigma}$ at low energies.  In our case its contribution to the
resistivity is as important as that of the quadratic part of
$H_{\sigma}$.  In the derivation below, we assume that the
inhomogeneities of the system are weak, e.g. $|u_{\rho}(x) -
u_{\rho}(0)| \ll u_{\rho}(0)$.

The resistivity can now be computed following a method similar to the one
outlined in Ref.~\onlinecite{Kostya} in the context of a quantum wire in
the Wigner crystal regime.  As one applies an electric current $I=I_0 \cos
\omega t$, the electrons start moving in the wire.  In the dc limit
$\omega \to 0$, we can assume that all electrons move in phase, so that at
time $t$ their position has shifted by a distance proportional to the
injected charge $q(t)=I_0 \omega^{-1} \sin \omega t$.  As a consequence,
we need to evaluate all the position-dependent parameters in the
Hamiltonian (\ref{bosonized_Hamiltonian}) at the true time-dependent
position of the electrons, which amounts to replacing $x \to x + q(t)/e
n(x)$.  In the regime of linear response, we only need to expand the
Hamiltonian to first order in $q(t)$,
\begin{equation}
H= \int dx \left( {\cal H}(x) + \frac{q(t)}{e n(x)} {\cal
H}^{\prime}(x) \right), \label{exp-ham}
\end{equation}
where ${\cal H}(x)$ is the Hamiltonian density when no current is
applied, and ${\cal H}^{\prime}(x)$ is obtained from ${\cal H}(x)$ by
replacing the position-dependent parameters
$g_{\sigma}(x)/\alpha(x)^2$, $u_{\nu} (x) K_{\nu} (x)$ and
$u_{\nu}(x)/K_{\nu}(x)$ (where $\nu=\rho,\sigma$) by their derivatives
with respect to $x$.

In the conventional Luttinger-liquid theory, the current $I=\dot{q}$ is
usually viewed as an excitation of the charge mode, and $q(t)$ thus
appears as a dynamical variable proportional to $\phi_{\rho}$. Then the
linear in $q$ part of Eq.~(\ref{exp-ham}) corresponds to cubic terms such
as $\phi_{\rho} (\partial_x \phi_{\rho})^2$. These cubic terms are usually
disregarded as irrelevant perturbations to the Luttinger liquid
Hamiltonian.  Nevertheless, the effect of such perturbations should be
addressed, because without them no contribution to transport arises from a
non-uniform interaction \cite{FLleads}.  In what follows, it will be more
convenient to treat $q(t)$ as an external parameter.

The oscillatory perturbation in the Hamiltonian (\ref{exp-ham}) acts as an
external driving force, which leads to the creation of spin and charge
excitations and dissipation of the energy from the driving force to the
wire. The energy $W$ dissipated into these excitations in unit time may be
obtained using the Fermi golden rule. In the limit of weak applied
current, it is expected to be quadratic in the amplitude $I_0$ of the
current oscillations.  This allows us, by comparison with the Joule heat
law $W=I_0^2 R/2$, to deduce the expression for the resistance $R$ of the
wire. Then in the dc limit $\omega\to0$ we find
\begin{equation} \label{dissipation}
R = \frac{i}{e^2 \hbar} \iint \frac{dx\,dy}{n(x)n(y)} \int
dt~ t\, \langle {\cal H}^{\prime}(x,t) {\cal H}^{\prime}(y,0)
\rangle,
\end{equation}
where $\langle \ldots \rangle$ corresponds to the thermodynamic
average.  The last integral in Eq.~(\ref{dissipation}) falls off
rapidly at $|x-y| \gg \hbar u_{\rho,\sigma}/T$.  Thus at $T \gg \hbar
u_{\rho,\sigma}/d$, we can reduce the expression (\ref{dissipation})
to a single integral in space, whose integrand we identify with the
resistivity $\rho(x)$ of the wire.

As both the charge and spin modes dissipate energy throughout the wire,
the total resistivity is given by the sum of a charge and a spin
contribution, $\rho(x)=\rho_{\rho}(x)+\rho_{\sigma}(x)$, which can be
computed separately.  Substituting the charge Hamiltonian
(\ref{bos-charge}) in Eq.~(\ref{dissipation}), and performing the
remaining time integral, one can extract the charge contribution to the
resistivity
\begin{equation} \label{charge}
\rho_{\rho}(x)=\frac{h}{8 \pi e^2} 
               \frac{T}{\hbar u_{\rho}(x)[n(x)]^2} 
               \left(\frac{\partial_x K_{\rho}(x)}{K_{\rho}(x)}\right)^2.
\end{equation}
This result holds at temperatures in the range $\hbar u_{\rho}/d \ll T \ll
D_{\rho}$, where the charge bandwidth $D_{\rho} \sim \hbar n u_{\rho}$.

One can use the expression (\ref{charge}) to recover our earlier result
(\ref{rhoresult}) for weakly interacting spinless electrons.  In this
case, upon bosonization the Hamiltonian of the system takes a form
equivalent to $H_{\rho}$ with $u_{\rho}(x) \to v_F$ and $K_{\rho}(x) \to 1
- \left[ V(0)-V(2k_F)\right] \eta(x)/2 \pi \hbar v_F$.  Substituting these
expressions into Eq.~(\ref{charge}) and expanding to second order in the
interaction, one reproduces the result (\ref{rhoresult}).

The spin contribution to the resistivity consists of two terms arising
from substituting the quadratic part and the cosine part of the spin
Hamiltonian in Eq.~(\ref{dissipation}).  The contribution of the quadratic
part of $H_{\sigma}$ can be obtained from Eq.~(\ref{charge}) by replacing
the charge parameters with their spin counterparts.  This result is
further simplified by using the low-energy expansion
$K_{\sigma}(x)=1+y_{\sigma}(x)/2$, where we introduced the dimensionless
parameter $y_{\sigma}(x) = g_{\sigma}(x)/\pi \hbar u_{\sigma}(x)$.

In the case of weakly interacting electrons, the correction to the
parameter $K_\sigma$ in the quadratic part of Eq.~(\ref{bosonized})
accounts for the $g_1$-coupling of the $z$ components of the electron
spins, while the cosine term in $H_{\sigma}$ is associated with the
remaining $x$ and $y$ components \cite{LL-th}.  Then from the spin
symmetry of the system, the cosine term of the spin Hamiltonian should
contribute twice as much to the resistivity as the quadratic part. We
expect this result to hold for arbitrarily strong interactions; in the
case of weakly interacting electrons, this conclusion can easily be
verified \cite{unpublished}. Combining these two terms, the spin
contribution to the resistivity reads:
\begin{equation} \label{spin}
\rho_{\sigma} (x) =  \frac{3h}{32\pi e^2}\frac{T}{\hbar u_{\sigma}(x)[n(x)]^2 }[\partial_x y_{\sigma} (x)]^2. 
\end{equation}
Again, we restricted ourselves to the range of moderately low
temperatures, $\hbar u_{\sigma}/d \ll T \ll D_{\sigma}$, where the spin
bandwidth is given by $D_{\sigma} \sim \hbar n u_{\sigma}$.

The comparison of the two contributions (\ref{charge}) and (\ref{spin}) to
the resistivity of the wire suggests the strongest effect in the regime of
low electron density, when the electron correlations are strong.  In this
case the exchange coupling $J$ of electron spins, which sets the spin
bandwidth $D_{\sigma}$, is strongly suppressed, so that $D_{\sigma} \ll
D_{\rho}$.  As a result, we expect the spin part (\ref{spin}) of the
resistivity to be the dominant contribution in this regime, due to the
reduced spin velocity.  It is worth pointing out that $y_{\sigma}$ is only
marginally irrelevant, so that while it renormalizes to zero at low
temperature, it does so logarithmically, as $1/ \log (D_{\sigma}/T)$.
Furthermore, most experimental measurements are carried out at fixed
temperature, while varying the electron density in the wire.  In this
configuration, the logarithmic dependence of $y_{\sigma}$ suggests that
this parameter increases as the interaction in the wire becomes stronger.

Our results are relevant to experiments on wires longer than the
length $l_{eq}$ associated with the processes of equilibration in the
moving frame.  In shorter wires, with length $L\ll l_{eq}$, we expect
the resistivity to be suppressed by an additional factor of order
$L/l_{eq}$.  This raises a fundamental question of the equilibration
in a one-dimensional system of interacting electrons.  In the weakly
interacting case, it is believed that the leading equilibration
mechanism is due to three-particle collisions \cite{3electron}
involving states near the bottom of the electronic band.  One expects
such processes to be strongly suppressed at low temperatures,
corresponding to a large $l_{eq}$.  However, such a treatment is not
applicable beyond the limit of weak interactions.  While we expect
stronger interactions to make thermalization easier, a detailed
investigation of the equilibration processes will be necessary to
access the full temperature dependence of the resistivity in this
regime.

In summary, we have shown that the interactions between electrons in a
long inhomogeneous quantum wire give rise to a finite resistivity
$\rho=\rho_\rho +\rho_\sigma$, given by Eqs.~(\ref{charge}) and
(\ref{spin}).  This resistivity is due to the weak violation of the
momentum conservation in electron-electron collisions, caused by the
inhomogeneities on long spatial scales $d\gg k_F^{-1}$.  Our results can
be tested experimentally by measuring the temperature and density
dependences of the resistance of long quantum wires.

We are grateful to A. V. Andreev, T. Giamarchi, and L. I. Glazman for
helpful discussions.  This work was supported by the U.S. Department of
Energy, Office of Science, under Contract No. DE-AC02-06CH11357.

\end{document}